# Personalized modeling for real-time pressure ulcer prevention in sitting posture

Vincent Luboz[1], Mathieu Bailet[2], Christelle Boichon Grivot[3], Michel Rochette[3], Bruno Diot[4], Marek Bucki[1], Yohan Payan[2]

[1]TexiSense, Montceau-les-Mines, France, {vincent.luboz, marek.bucki}@texisense.com;

[2] Univ. Grenoble Alpes, CNRS, Grenoble INP, TIMC-IMAG, F38000 Grenoble France, {Mathieu.Bailet, Yohan.Payan}@univ-grenoble-alpes.fr;

[3]ANSYS, Lyon, France, {michel.rochette, christelle.boichon}@ansys.com;

[4]IDS, Montceau-les-Mines, France, b.diot@ids-assistance.com;

Corresponding author

Yohan Payan

Yohan.Payan@univ-grenoble-alpes.fr

TIMC-IMAG Laboratory
Faculté de Médecine
Pavillon Taillefer
38706 La Tronche cedex
France

Tel: +33 (0)4 56 52 00 01 - Fax: +33 (0)4 56 52 00 55

Abstract,
Ischial pressure ulcer is an important risk for every paraplegic person and a major public health issue. Pressure ulcers appear following excessive compression of buttock's soft tissues by bony structures, and particularly in ischial and sacral bones. Current prevention techniques are mainly based on daily skin inspection to spot red patches or injuries. Nevertheless, most pressure ulcers occur internally and are difficult to detect early. Estimating internal strains within soft tissues could help to evaluate the risk of pressure ulcer. A subject-specific biomechanical model could be used to assess internal strains from measured skin surface pressures. However, a realistic 3D non-linear Finite Element buttock model, with different layers of tissue materials for skin, fat and muscles, requires somewhere between minutes and hours to compute, therefore forbidding its use in a real-time daily prevention context. In this article, we propose to optimize these computations by using a reduced order modeling technique (ROM) based on proper orthogonal decompositions of the pressure and strain fields coupled with a machine learning method. ROM allows strains to be evaluated inside the model interactively (i.e. in less than a second) for any pressure field measured below the buttocks. In our case, with only 19 modes of variation of pressure patterns, an error divergence of one percent is observed compared to the full scale simulation for evaluating the strain field. This reduced model could therefore be the first step towards interactive pressure ulcer prevention in a daily set-up.

Highlights
- Buttocks biomechanical modelling,
- Reduced order model,
- Daily pressure ulcer prevention.

1. Introduction

More than 50,000 people suffer from paraplegia in France (HAS, 2009). Among them, 80 % will develop a pressure ulcer (PU). PUs result from the compression of soft tissues between a bony prominence and a supporting surface, e.g. a wheelchair. PUs start near bony structures and progress towards the skin, causing significant subcutaneous damages before being clinically observable. Paraplegic persons use cushions to evenly distribute pressure below their buttocks and try to regularly change their sitting posture to prevent PUs. Unfortunately, this is not sufficient since subjects' attention can decrease over time.
Measuring pressures at the skin/seat interface can help to prevent skin injuries (Pipkin and Sprigle, 2008), but these measurements cannot predict dangerous internal tissue strains responsible for deep PUs (Linder-Ganz et al., 2008). Intuitively, a person having sharper ischial tuberosities is more at risk of developing a PU than a person with blunt ischia, even with similar pressures below their buttocks (Sopher et al., 2010). Estimating internal strains is consequently the only solution to assess PU risk level. Animal studies (Loerakker et al., 2011) suggest that two strain thresholds should be monitored as there is a long term risk of ischemic PU if the internal strains are above 20 % for about 2 hours, and a short term risk of mechanical PU if the strains exceed 50 % for about 10 minutes (these values, namely 20% for 2 hours and 50% for 10mn, are orders of magnitudes that should be taken with cautious in the case of human tissues). These internal strains can only be estimated with a subject-specific biomechanical model to study mechanical interactions between soft tissues and bony prominences (Elsner and Gefen, 2008).

The gluteal region has been modelled by several groups for different applications. A real-time 2D biomechanical Finite Element (FE) model using a linear elastic constitutive law for the muscles (E = 8.5 kPa, $\upsilon$ = 0.49) and the fat/skin tissues (E = 32 kPa, $\upsilon$ = 0.49) was proposed by (Linder-Ganz et al., 2009) to evaluate the internal strains in the buttocks of a paraplegic person. Earlier, a 3D FE hyper-elastic model using a Neo Hookean constitutive law to simulate the skin (E = 150 kPa, $\upsilon$ = 0.46) and a Mooney Rivlin constitutive law to simulate the other soft tissues (A1 = 1.65 kPa, A2 = 3.35 kPa, $\upsilon$ = 0.49) was proposed by (Verver et al., 2004). It showed that the pressure distribution and therefore the internal strains depend on the stiffness of the chair cushion, on the constitutive material of the buttocks' soft tissues, and on the subject's posture. Using another 3D FE hyper-elastic model, (Luboz et al., 2014) have studied the simulation sensitivity to the elasticities of the various layers of soft tissues in the buttocks, concluding that generic values of 200 kPa, 30 kPa, and 100 kPa can be used for the skin, fat and muscles, in a Neo Hookean context.

The goal of this paper is to study the feasibility of coupling a 3D biomechanical model with an embedded pressure mat to form a personalized PU risk assessment device. To this aim, we introduce a reduced order modeling technique to run our FE hyper-elastic buttock model in real-time and to assess internal strains over time in different sitting postures.

2. Materials and Methods

2.1. Buttock FE hyper-elastic model

The modeling and simulations are performed within the ArtiSynth open source framework (Lloyd et al., 2012) (www.artisynth.org) using a dynamic numerical solver. This study is based on the morphology of a male subject (38 years old, 100 Kg and 1.90 m). The external surfaces of the skin, muscles and bones were semi-automatically segmented from a CT scan (image size 512x512x403, resolution 0.97x0.97x1 mm$^3$), using the ITK-Snap snake tool (Yushkevich et al., 2006). The muscles were segmented as a single entity as they were too difficult to separate. To have the least deformed contours of the modeling domain, the subject was lying on his side in the CT scan, which limited the distortion in one side of his buttocks. However, deformations induced by gravity were not compensated for. The external surfaces of our model are a combination of the least deformed side and its mirrored, contralateral part.

A meshing tool provided by Texisense Company was used to generate a three layer FE mesh, corresponding to the skin, fat and muscles. The hex-dominant mesh is composed of 27,649 linear elements, fig. 1a. The skin is a 1-element 1.5mm-thick layer of elements (Hendriks et al., 2006) at the FE mesh surface. The segmented muscle surface delimits the muscle elements. Finally, the elements between the skin and muscle layers are considered as fat tissues. Fig 1b shows a mesh cross section after identifying these structures. In this model, the bones are assumed to be rigid. The nodes at the tissue/bone interface are attached to these bones with no sliding.

The three soft tissues layers are modeled using a Neo Hookean constitutive material (Bonnet & Wood, 1997) to simulate a nonlinear hyper-elastic behavior. The strain energy density function W is given by:

$$W=C_{10}(I_1-3)+(J-1)^2/D$$

Where $I_1$ is the first invariant of the left Cauchy-Green deformation tensor, $C_{10}$ is a material parameter (under a small strain hypothesis, it is related to the Young's modulus $E \approx 6 * C_{10}$), and J is the determinant of the deformation gradient F. D is the material incompressibility, related to the Poisson ratio $\upsilon$ [$D=(1-2\upsilon)/C_{10}$].

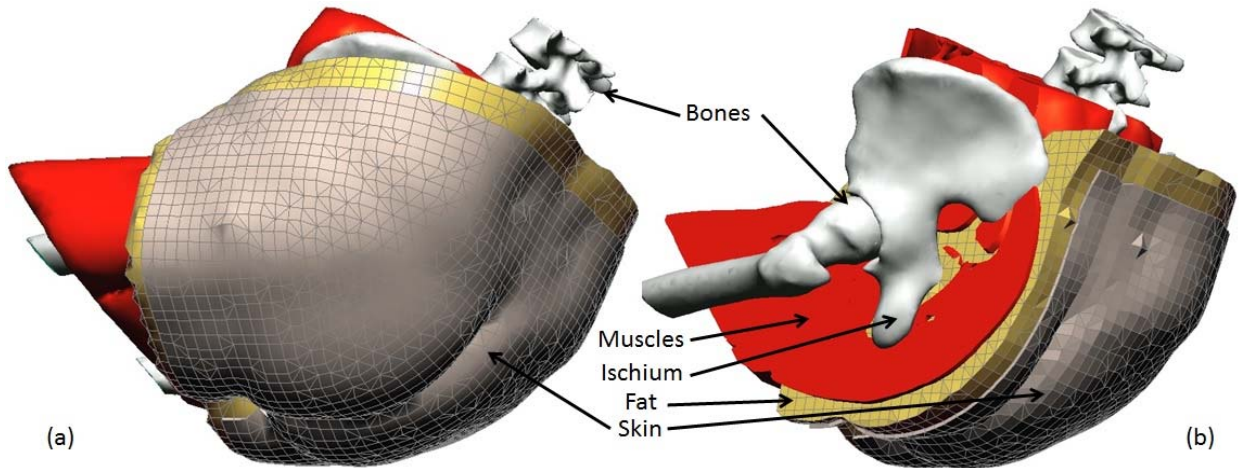

Fig. 1: (a) Finite element model of the buttocks, (b) Sagittal cross section showing the three layers of materials defining the buttock model: skin, fat and muscles. Bones are assumed to be rigid and fixed to the nearby FE nodes.

The mechanical properties of the different soft tissue layers are taken from (Luboz et al., 2014). Equivalent Young modulus values of 200 kPa ($C_{10}$ = 33 kPa), 30 kPa ($C_{10}$ = 5 kPa), and 100 kPa ($C_{10}$ = 17 kPa) are respectively chosen for the skin, fat and muscles. A Poisson ratio of 0.49 is used for these three layers, as they are assumed to be quasi-incompressible. These mechanical properties are generic: for better accuracy, subject specific properties would need to be measured (for example, using an elastography procedure or an *in vivo* aspiration device (Schiavone et al., 2010)).

2.2. Boundary conditions

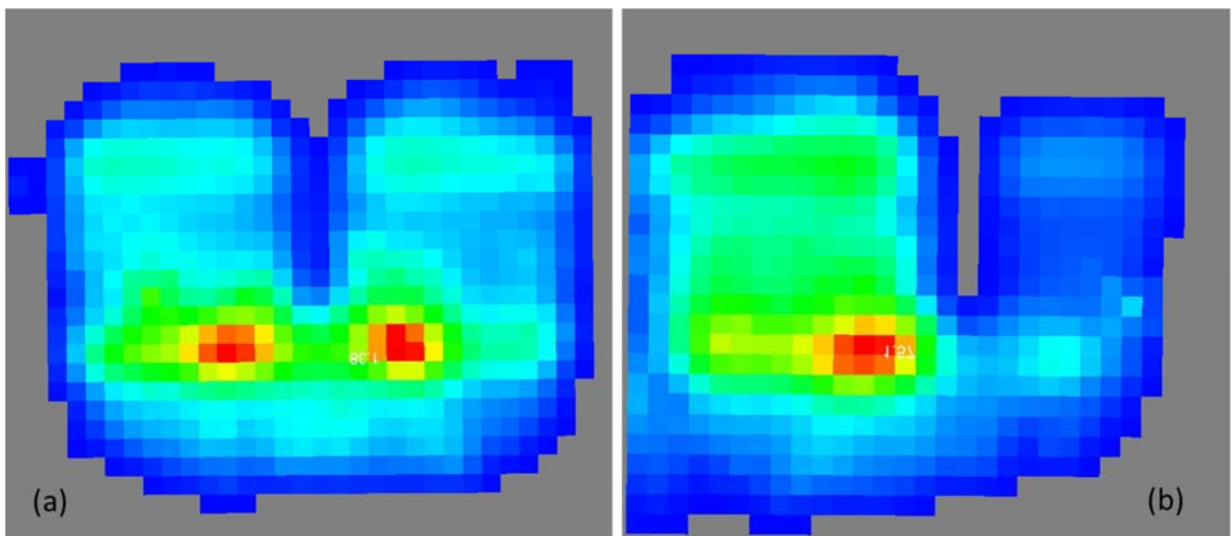

Fig. 2: Two pressure map examples where the subject sits with (a) his arms on his knees, back away from the back-rest, and (b) most of his weight on his left buttock. The highest measured pressure, in red, is (a) 1.38 N.cm$^{-2}$ and (b) 1.57 N.cm$^{-2}$.

The acquisition of pressures below the sitting subject is performed using a commercial pressure sensor: TexiMat (www.texisense.com). It is composed of a textile matrix of 32x32 sensors of 1.5cmx1.5cm each. Pressure acquisition lasted three hours, during which the subject could move freely, fig. 2: put his back on the chair rest, lift his feet, move his weight on either side of the buttocks. Shear loads on the cushion were not measured by the sensor and the subject was asked to avoid as much as possible leaning postures. To determine where to apply the pressure on the model, the ischial tuberosities are detected on the pressure frames. The ischial tuberosity footprints on the frames are two relatively close pressure peaks. The first stage of the detection pipeline consists in a Gaussian filtering to remove sensor noise (fig. 3a). In the second stage, a Laplacian filter with a 9-pixel kernel followed by a binary threshold on the normalized resulting image are applied to detect blob-like structures that could correspond to ischial pressure footprint (fig. 3b). This filter is independent of pressure magnitudes. Once the pressure image has been thresholded, several pairs of regions of interest can potentially correspond to the ischia. At the last stage, we use a geometric filter based on the relative position of each pair of centroids to find the most appropriate one (fig. 3c). The retained ischia are matched below those of the FE model. Finally, the buttock's surface FE nodes are projected orthogonally onto the pressure array. A bilinear interpolation between the four sensors surrounding each projected node is performed to compute the pressure applied at each FE node and to produce a continuous 3D pressure field. The normal at the buttock surface is taken as the pressure normal. In order to ensure simulation convergence, interface pressures are applied following a linear ramp from 0 % at 0.1 s to 100 % at 1.1 s. After 1.1 s, the pressure is maintained until the continuum reaches a steady state at about 1.3 s.

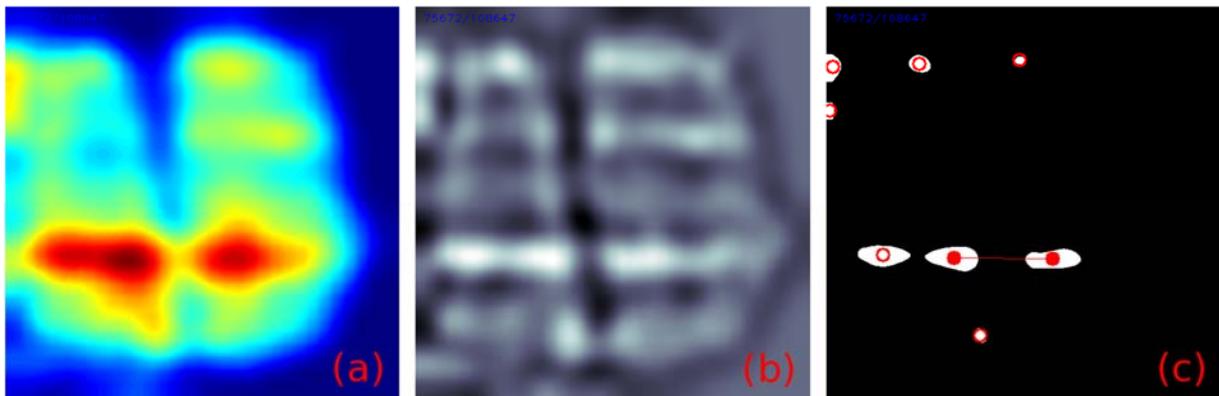

Fig 3: Ischial detection with (a) Gaussian smoothing, (b) Laplacian filtering, and (c) thresholding and morphological filtering to pair the ischium (empty circles are possible matches, full circles are the detected ischium).

2.3. Reduced Order Model

For each pressure map, the FE buttock model requires about 20 minutes to compute the simulation and assess internal strains. To design a solution that will provide a real-time computation of our FE model we used the Reduced Order Modelling (ROM) technique developed by ANSYS (www.ansys.com). For a given subject, ROM is applied on the outputs of the FE solver (ArtiSynth) which computes the model deformations. Note that for each new subject, a new ROM must be computed as subjects' morphology and tissue mechanical properties vary greatly and thus influence internal strains.
More precisely, an off-line learning step is first performed for each subject model: a set of pressure frames acquired for the subject is compressed into a small number of modes by using a proper orthogonal decomposition inspired by (Chaturantabut et al., 2012). This technique allows each pressure frame to be linearly described by a small set of scalar coefficients. Then the most relevant pressure frames to describe the pressure variations are selected, based on an optimal repartition of the learning points in the mode coefficients' space. The FE analyses corresponding to this training set of typical boundary conditions are carried out and the resulting strain fields are also compressed into a small number of modes. A machine-learning approach is then used to learn the relationship between the strain field modes and the pressure map modes. This approach uses response surface techniques based on Kriging interpolation and available in ANSYS 18.0 Design Xplorer.

Once built, the ROM interactively gives an accurate approximation of the solver solution for a new set of interface pressures applied on the model. For this, the new pressure map is first projected by the ROM algorithm onto the basis of pressure modes. Then, from the projection coefficients, the ROM evaluates the response surface that gives the strain field coefficients. The whole strain field is finally built as a linear combination of strain modes. All these steps are performed in nearly real-time.

The accuracy of this ROM process is measured on the resulting strain fields and it depends on three parameters: the number of pressure modes chosen to compress the pressure field, the targeted precision of the statistical reduction of the strain field (related to the number of modes used to represent the strain field), and the number of pressure frames in the training set.

3. Results

In our case, the ROM was generated based on 9,100 pressure frames selected from a 3-hour recording acquired at a framerate of 1 Hz below the sitting subject. Statistical analysis showed that 19 pressure modes were able to describe the 9,100 pressure frames with a root mean square error of 1.98 %. Then, the most significant pressure frames were selected based on their variations across those modes, to serve as a training set for the ROM. We chose three training sets: the first 100, 150, and 200 most significant frames. We then selected 50 random frames, outside these training sets, to serve as validation frames for the comparison between the strains estimated by the ROM and the strains computed by the actual FE model.

A sensitivity analysis was performed to assess the three ROM parameters' influence onto the accuracy of the estimated strain field.

The strain field error on each validation pressure frame $f$ was calculated by using the Euclidean norm of the error field, given by the node to node ($i = 1$ to $n$, with $n$ the number of FE nodes) difference between the $m = 50$ exact strain fields at node $i$ in frame $f$, *FEstrain(f, i)*, and the $m$ estimated strain fields *ROMstrain(f, i)*. This error was then normalized with the mean Euclidean norm value of those exact strain fields. Finally we considered the mean error over the mean value, *MeanEMV*, and the max error over the mean value, *MaxEMV*, of this relative error on the 50 validation frames to assess the accuracy of the estimated strain field:

$$MeanEMV = \frac{1}{m}\sum_{f=1}^{m} \frac{\sqrt{\sum_{i=1}^{n}(FEstrain(f,i) - ROMstrain(f,i))^2}}{\frac{1}{m}\sum_{f=1}^{m}\sqrt{\sum_{i=1}^{n}FEstrain(f,i)^2}} \quad (1)$$

$$MaxEMV = \frac{\max\{f=1...m\}\sqrt{\sum_{i=1}^{n}(FEstrain(f,i) - ROMstrain(f,i))^2}}{\frac{1}{m}\sum_{f=1}^{m}\sqrt{\sum_{i=1}^{n}FEstrain(f,i)^2}} \quad (2)$$

At first, we fixed the training set frame number to 100, and estimated the sensibility over the two other parameters, namely the number of pressure modes and the targeted precision. Fig. 4 summarizes the sensitivity analysis over MeanEMV for 5, 10, 15 and 19 pressure modes, and a targeted mean precision of the statistical reduction for the strain field of 0.2, 0.5, 1, 2, and 5 %. This figure shows that a plateau is reached if a targeted precision parameter of 0.5 % and 19 modes are used, with an actual MeanEMV of 1.09 %. Preferring 15 modes could be interesting when ported to a micro-controller as it reduces the computation time with only a slight increase of the errors compared to 19 modes.

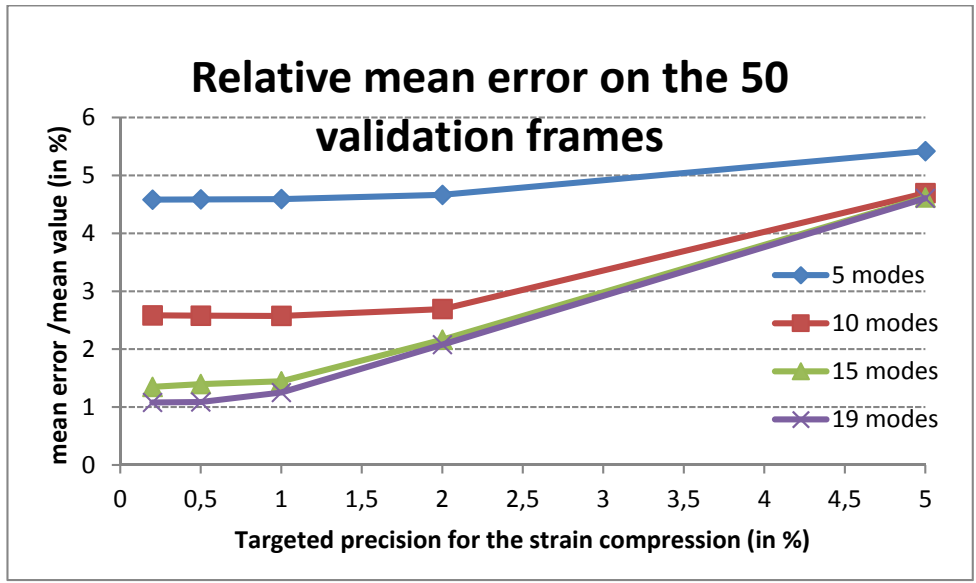

Fig. 4: Mean Error over Mean Value (MeanEMV) for 5, 10, 15 and 19 pressure modes, and a targeted precision for the strain statistical reduction of 0.2, 0.5, 1, 2, and 5 %.

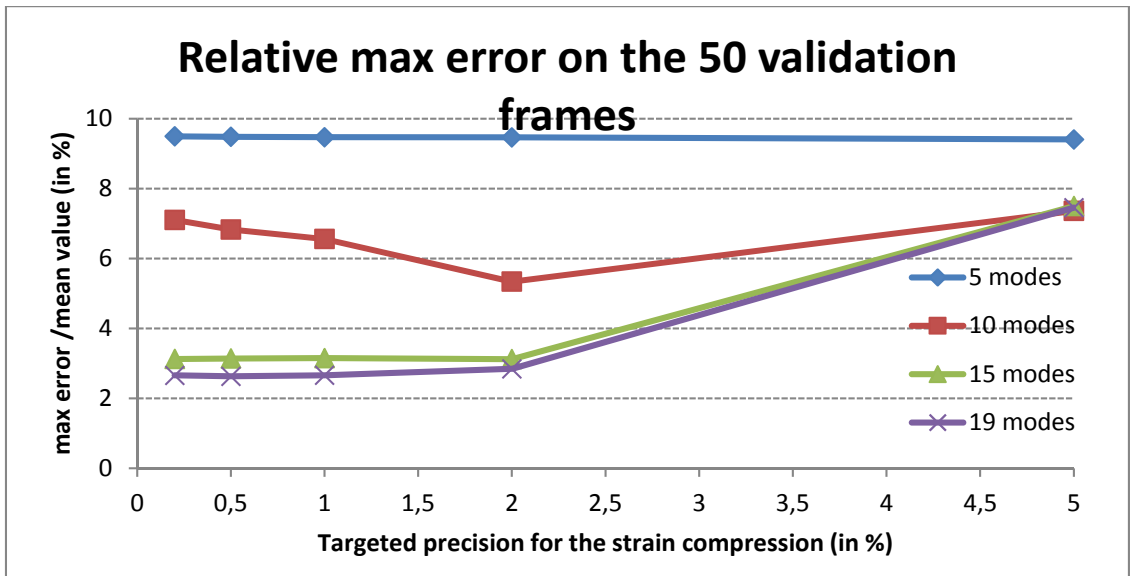

Fig. 5: Max Error over Mean Value (MaxEMV) for 5, 10, 15 and 19 modes, and a targeted precision of 0.2, 0.5, 1, 2, and 5 %.

The same conclusions can be drawn from Fig. 5, summarizing the sensitivity analysis over MaxEMV. A plateau is reached if a targeted precision parameter of 2 % and 19 modes are used, with an actual MaxEMV of 2.63 %. In this figure, outliers appear on the 10-mode curve for targeted precisions below 1 %. This is due to a specific node in the buttock model, presenting a large strain, not representative of the rest of the FE mesh and leading to a strain misevaluation for a validation pressure frame. This event can be related to the uncertainties induced by considering maximal strains for PU prevention. Using "cluster analysis" could overcome this limitation (see Bucki et al., 2016 for details) since it is based on the study of the largest volume of contiguous elements having a strain above a certain threshold, therefore smoothing possible misinterpretation due to a single node with a high strain value.

We kept 19 pressure modes and a targeted precision of the strain field statistical reduction of 0.5 % for our second sensitivity analysis aiming at defining an optimized number of learning frames over the three sets defined earlier. We analyzed the MeanEMV and MaxEMV ratios given by the ROM on the 50 random validation frames, for 100, 150 and 200 learning frames. Fig. 6 shows that a plateau is reached for 150 learning frames with a MaxEMV of 1.64 %. A ratio of 0.79 % is recorded for MeanEMV for 200 frames.

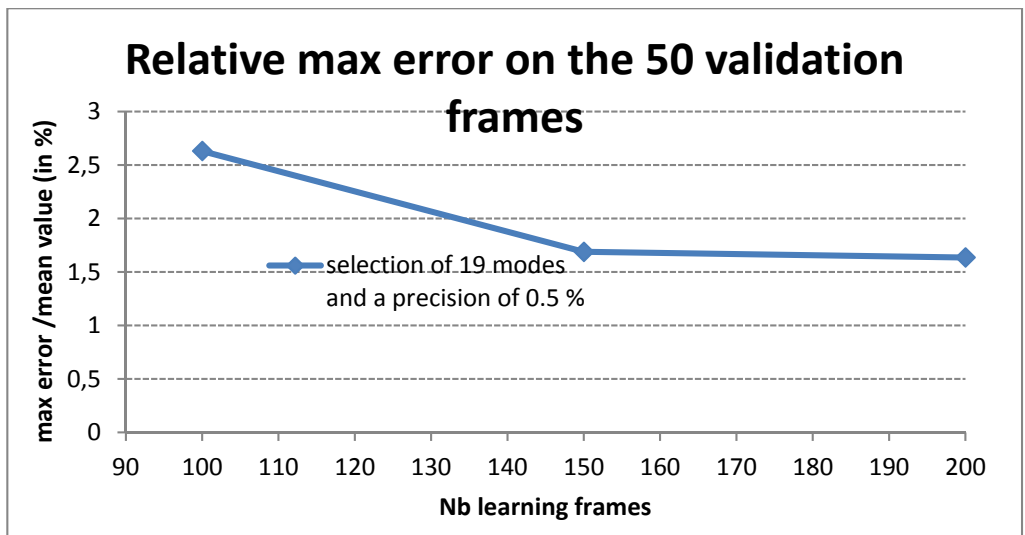

Fig. 6: Max Error over Mean Value (MaxEMV) for 100, 150 and 200 learning pressures frames, 19 modes, and a 0.5 % targeted precision.

With no more than 200 pressure frames in the learning set up, 19 pressure modes and a targeted precision of 0.5 % for the strain compression, it is therefore possible to reach accuracies of 0.79 % and 1.64 % respectively for MeanEMV and MaxEMV. Using these ROM settings, it is thus possible to compute accurately the strain field for any acquired pressure frame in less than a second, on a standard PC, instead of 20 minutes for the FE hyper-elastic model. The resulting deformation field, fig. 7, permits a mapping of the PU formation risk in the buttock soft tissues, while the pressures evolve below the subject.

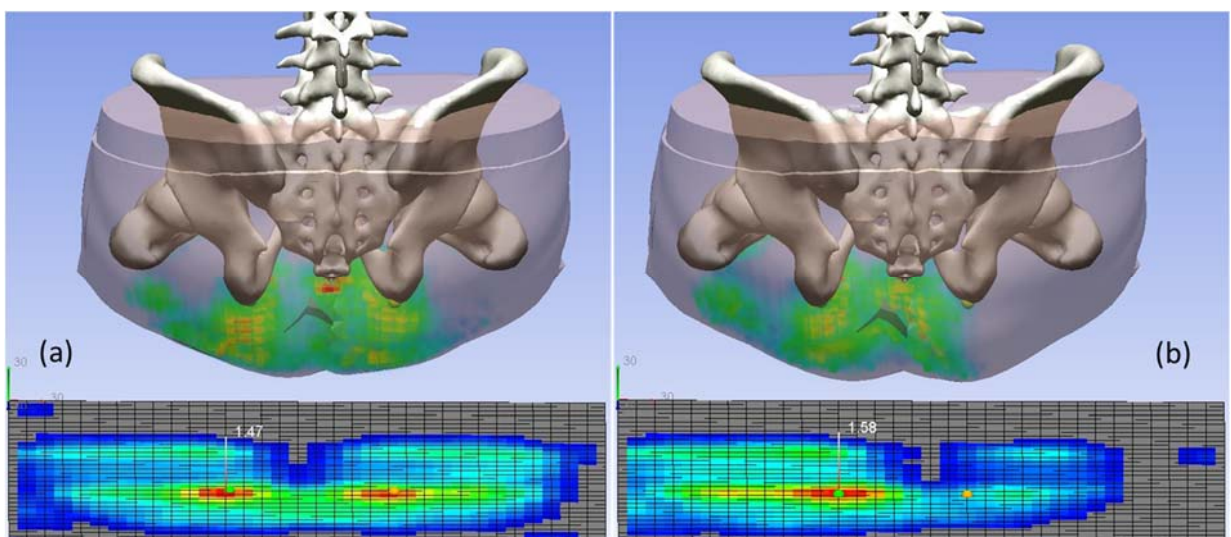

Fig. 7: Two examples of deformation fields, where the subject sits with: (a) his arms on his knees, back away from the backrest, leading to an even deformation field in his soft tissues below the ischial tuberosities and sacrum (maximum strain in red is 23 %), and (b) most of his weight on his left buttock, leading to higher strains below his left ischia (maximum strain is 31 %).

4. Conclusions

A 3D FE non-linear subject-specific buttock model was generated to provide pressure ulcer prevention in the gluteal region. The personalized FE model is composed of three soft tissue layers: skin, fat and muscles, all modelled with Neo-Hookean constitutive materials. PU prevention can be achieved by continuously monitoring strains within soft tissues undergoing compression. Because FE estimation of internal strains on such a complex model takes about 20 minutes, a Reduced Order Model (ROM) was introduced. The ROM was built by statistically reducing 9,100 pressure frames acquired below the sitting subject, and by projecting the responses of the personalized biomechanical model on 19 pressure modes. A sensitivity analysis showed that the optimal ROM settings were 19 modes, a targeted precision of the strain field statistical reduction of 0.5 %, and 200 learning frames. Using these parameters, the optimal ROM provides strain fields in less than a second, with an accuracy of 0.79 %.

To demonstrate the feasibility of a PU prevention device, we used the TexiMat textile pressure sensor to continuously measure pressure frames under the sitting subject. These measurements, coupled with real-time strain evaluation inside his gluteal soft tissues provided by the ROM, and compared to the time and strain threshold proposed by (Loerakker et al., 2011), allowed interactive PU risk assessment. Since all on-line computations can be ported to a micro-controller embedded within a pressure mat placed on a wheelchair, the ROM could be used in a daily PU prevention set-up. The resulting device could wirelessly send warnings, through a smartphone or a watch, to the wheelchair user or the nursing staff in a clinical setting. The training phase would take less than an hour, while the patient would assume various key positions on his chair (patient on one side of his buttocks, on both sides, with his back on the chair…), to initialize the ROM with the measured pressures under his buttocks, and consequently avoiding a risk of creating a pressure ulcer for patients at risk.

Nevertheless, before using this reduced model in a daily PU prevention set up, the software needs some adjustments. Besides porting the existing ROM technology onto a mobile platform, another technical hurdle is the ability to easily generate an accurate subject-specific biomechanical model from available medical data. To this aim, the method proposed by (Bucki et al., 2016) could be of interest and would ease the generation of the patient specific reduced model needed for each new subject.

**Acknowledgements**

Competing interests: Some authors are involved with TexiSense (http://www.texisense.com/home_en) and some with ANSYS (http://www.ansys.com/).

Funding: This work was partly funded by the 2010 ANR TecSan IDS project, by the CAMI Labex (ANR-11-LABX-0004).

Ethical approval: the study was approved by the French institutional review board of Grenoble (IRB 5891 (CECIC) for Rhône-Alpes-Auvergne).

**References**
1. Bonet, J., and Wood, R.D., 2008. Nonlinear Continuum Mechanics for Finite Element Analysis. Cambridge University Press.
2. Bucki M., Luboz V., Perrier A., Champion E., Diot B., Vuillerme N., & Payan Y., 2016. Clinical workflow for personalized foot pressure ulcer prevention. Medical Engineering and Physics, 38(9): 845–853.
3. Chaturantabut, S., & Sorensen, D. C., 2012. A state space error estimate for POD-DEIM nonlinear model reduction. SIAM Journal on numerical analysis, 50(1), 46-63.
4. Elsner, J.J., and Gefen, A., 2008. Is obesity a risk factor for deep tissue injury in patients with spinal cord injury? Journal of Biomechanics, 41:3322–3331.

5. HAS: Haute Autorité de Santé, Paraplégie, 2009.
6. Hendriks, F.M., Brokken, D., Oomens, C.W.J., Bader, D.L., and Baaijens, F.P.T., 2006. The relative contributions of different skin layers to the mechanical behavior of human skin in vivo using suction experiments. Medical Engineering & Physics, 28:259–266.
7. Linder-Ganz, E., Shabshin, N., Itzchak, Y., Yizhar, Z., Siev-Ner, I., and Gefen, A., 2008. Strains and stresses in sub-dermal tissues of the buttocks are greater in paraplegics than in healthy during sitting. Journal of Biomechanics, 41:567–580.
8. Linder-Ganz, E., Yarnitzky, G., Yizhar, Z., Siev-Ner, I., and Gefen, A., 2009. Real-Time Finite Element Monitoring of Sub-Dermal Tissue Stresses in Individuals with Spinal Cord Injury: Toward Prevention of Pressure Ulcers, Annals of Biomedical Engineering, 37(2):387–400.
9. Loerakker, S., Manders, E., Strijkers, G.J., Nicolay, K., Baaijens, F.P.T., Bader, D.L., Oomens, C.W.J., 2011. The effects of deformation, ischaemia and reperfusion on the development of muscle damage during prolonged loading. Journal of Applied Physiology, 111(4): 1168-1177.
10. Lloyd, J.E., Stavness, I., and Fels, S., 2012. Artisynth: a fast interactive biomechanical modeling toolkit combining multibody and finite element simulation. In: Payan Y. (Ed.), Soft Tissue Biomechanical Modeling for Computer Assisted Surgery, Studies in Mechanobiology, Tissue Engineering and Biomaterials, 11:355–394.
11. Luboz V., Petrizelli M., Bucki M., Diot B., Vuillerme N. & Payan Y., 2014. Biomechanical Modeling to Prevent Ischial Pressure Ulcers. Journal of Biomechanics, 47 (2014) 2231-2236.
12. Pipkin, L., and Sprigle, S., 2008. Effect of model design, cushion construction, and interface pressure mats on interface pressure and immersion. Journal of Rehabilitation Research & Development, 45:875–882.
13. Schiavone P., Promayon E. & Payan Y. (2010). LASTIC: a Light Aspiration device for in vivo Soft TIssue Characterization. Lecture Notes in Computer Science, Vol. 5958, pp. 1-10.
14. Sopher, R, Nixon, J., Gorecki, C., and Gefen, A., 2010. Exposure to internal muscle tissue loads under the ischial tuberosities during sitting is elevated at abnormally high or low body mass indices. Journal of Biomechanics, 43:280–286.
15. Verver, M.M., van Hoof, J., Oomens, C.W.J., Wismans, J.S.H.M. and Baaijens, F.P.T., 2004. A Finite Element Model of the Human Buttocks for Prediction of Seat Pressure Distributions, Computer Methods in Biomechanics and Biomedical Engineering, 7(4):193-203.
16. Yushkevich, P.A., Piven, J., Hazlett, H.C., Gimpel Smith, R., Ho, S., Gee, J.C., and Gerig, G., 2006. User-guided 3D active contour segmentation of anatomical structures: Significantly improved efficiency and reliability. Neuroimage, 31(3):1116-28.